\documentclass[10pt,letterpaper]{article}
\usepackage{opex3,array,multirow,graphicx}
\newcolumntype{x}[1]{>{\centering\hspace{0pt}}p{#1}}

\newcommand{\ket}[1]{\mbox{\ensuremath{|#1\rangle}}}

\begin{document}




\title{Mode expansion and Bragg filtering for a high-fidelity fiber-based photon-pair source}

\author{Alexander Ling*, Jun Chen, Jingyun Fan and Alan Migdall}

\address{National Institute of Standards and Technology, Gaithersburg, MD 20899 \\ Joint Quantum Institute, University of Maryland, College Park, MD 20742}

\email{aling@nist.gov} 


\begin{abstract*}
We report the development of a fiber-based single-spatial-mode source of photon-pairs where the efficiency of extracting photon-pairs is improved over a previous source [18] through the use of fiber-end expansion and Bragg filters.
This improvement in efficiency enabled a spectrally bright and pure photon-pair source having a small second-order correlation function (0.03) and a raw spectral brightness of 44,700 pairs $\mbox{s}^{-1} \mbox{nm}^{-1} \mbox{mW}^{-1}$.
The source can be configured to generate entangled photon-pairs, characterized via optimal and minimal quantum state tomography, to have a fidelity of 97\% and tangle of 92\%, without subtracting any background.\\
\end{abstract*} 

\ocis{(270.4180) Multiphoton processes; (190.4370) Nonlinear optics, fibers; (190.4380) Nonlinear optics, four-wave mixing } 

\section{Introduction}

The contemporary workhorse method for obtaining photon-pairs has been Spontaneous Parametric Down Conversion (SPDC) \cite{coinc70} in nonlinear crystals.
Typical SPDC sources employ bulk crystals whose output is coupled into single-mode fibers \cite{kurtsiefer01,fed07}.
However, because SPDC emission from bulk crystals is inherently spatially multi-mode, only a fraction of the two-photon light can be collected into a single-mode fiber.
This is one reason that has prompted numerous studies on SPDC inside waveguides \cite{uren04,fiorentino07,chen09,christ09,karpinski09}, as the overlap of the total emitted light with a single spatial mode can be much greater.

Another approach, which is the subject of this paper, is to generate photon-pairs via Spontaneous Four-Wave Mixing (SFWM) inside optical fibers \cite{fiorentino02, rarity05, dyer09, fan05, takesue05, chen06, lee06}.  
Of particular interest is SFWM inside of a solid-core photonic-crystal fiber \cite{russell03} (PCF), which has spatial mode sizes typically an order of magnitude smaller than in conventional fibers, resulting in much higher nonlinearity.
Combined with appropriate phase-matching conditions, SFWM in PCFs has enabled very bright polarization-entangled photon-pair sources (even after aggressive spectral filtering) operating at room temperature \cite{fiberent07}.
However, the PCF source still has outstanding issues.
Most significantly, the performance of PCF sources suffer from low extraction efficiency;  losses are typically high when the pairs are to be prepared into a well-defined bandwidth with a useful single spatial-mode.

In this paper we report a novel method of improving the extraction efficiency, compared to earlier implementations of PCF sources \cite{goldschmidt08,fulconis07}.
This was enabled by incorporating novel (and simple-to-use) elements - end-tapered PCFs, high-efficiency reflection Bragg gratings and high-transmission band-pass filters.
Our implementation has enabled us to incorporate a higher count rate with higher purity as compared to other single-mode photon-pair SFWM sources \cite{goldschmidt08,fulconis07}.
The photon-pair purity of the source, determined using the second order correlation function ($\mbox{g}^{(2)}(0)$), was measured to be as small as 0.007.
When the source is used to generate polarization-entangled photon-pairs, the fidelity (to a Bell state) and tangle are measured to be 97\% and 92\%, respectively.
Indistinguishable photons have also been heralded by detection of their twins, which exhibit a high level of indistinguishability via the demonstration of a 82\% raw visibility Hong-Ou-Mandel interference dip.

\section{Photon-pair purity}
As the basic theory of SFWM inside fibers has been well described previously \cite{chen05}, we begin by describing our experimental parameters.

\subsection{Efficient extraction of photon-pairs.}
\begin{figure} \centerline{\scalebox{0.4} { \includegraphics{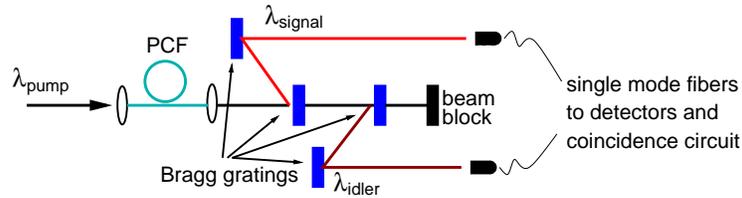}}} \caption{\label{fig:fiberscheme} (Color online) Layout of first experiment for pumping a PCF and collecting photon-pairs.  Photon-pairs are detected via a start-stop coincidence circuit.   Reflection Bragg gratings separate signal and idler from the pump.  Using two gratings per arm suppresses the pump light by up to 180 dB.  } \end{figure}
\begin{table}[t] 
\caption{ Extraction efficiencies for different PCF sources. Where possible we have provided the experimental uncertainties (1 standard deviation).  The values for $\eta_{fiber}$ is taken by assuming 4\% reflection loss at an uncoated glass surface. 
} 
\hspace{-0cm} \begin{tabular}{p{4cm}x{1.25cm}x{1.25cm}|x{0.9cm}x{0.8cm}|x{0.9cm}x{0.8cm}} 
\multicolumn{7}{c}{Efficiency (\%)}\tabularnewline \hline\hline 
\multirow{2}{*}{} & \multicolumn{2}{c|}{Our source} & \multicolumn{2}{c|}{Source: \cite{goldschmidt08}} & \multicolumn{2}{c}{Source: \cite{fulconis05}} 
\tabularnewline spectral selection method & \multicolumn{2}{c|}{Bragg Grating} & \multicolumn{2}{c|}{Monochromator} & \multicolumn{2}{c}{Etalon Filter} 
\tabularnewline & signal & idler & signal & idler & signal & idler 
\tabularnewline \hline $\eta_{fiber}$ (calc.) & 96 & 96 & 96 & 96 & \multicolumn{2}{c}{N.A.} 
\tabularnewline $\eta_{lens}$ (meas.)& $98 \pm 1$ & $98 \pm 1$& 75 & 70 & \multicolumn{2}{c}{N.A.} 
\tabularnewline $\eta_{spectral}$ (meas. \& calc.) & $28 \pm 1$  & $38 \pm 1$  & 16  & 27 & \multicolumn{2}{c}{N.A.}
\tabularnewline $\eta_{coupling}$ (meas.) & $50 \pm 1$ & $50 \pm 1$ & 53 & 58 & \multicolumn{2}{c}{60-65}\\ 
\hline $\eta_{signal}\mbox{, } \eta_{idler}$  & $13 \pm 0.6$ & $18\pm 0.6$ & 6.1 & 11 & 38 & 36
\tabularnewline 
$\eta_{pair}=\eta_{signal}\times\eta_{idler}$  & \multicolumn{2}{c|}{$2.3 \pm 0.1$}& \multicolumn{2}{c|}{0.7} & \multicolumn{2}{c}{14}
\tabularnewline $\eta_{det}$  & $56\pm0.6$ & $43\pm 0.4$ & 60 & 50  & 60 & 33 
\tabularnewline \hline Single Photon Detection Eff. & $7.3 \pm 0.3$ & $7.7 \pm 0.3$ & 3.7 & 5.5 & 23 & 12
\tabularnewline Photon-pair Detection Eff.  & \multicolumn{2}{c|}{$0.56 \pm 0.03$} & \multicolumn{2}{c|}{0.21} & \multicolumn{2}{c}{2.8}\\\hline\hline \end{tabular} \end{table}

In our photon-counting experiments (Fig. \ref{fig:fiberscheme}), we work with PCFs engineered to be polarization maintaining along two principal axes.  
To maximize the nonlinear gain, the pump polarization is aligned with the axis which exhibits higher nonlinearity; this configuration also enables the generation of co-polarized photon-pairs.

The PCF we use has a length of one meter and a nominal zero-dispersion wavelength of $745 \pm 5$ nm, and was obtained from Crystal Fibre A/S \cite{crystalfibre,disclaimer}. 
The core diameter of the PCF is $ \approx 2 \mu$m, but the fiber structure at the ends was collapsed over a length of 50 $\mu$m to yield a 15 $\mu$m mode resulting in a smaller divergence (NA$\approx0.3$, from manufacturer's datasheet) compared to the non-tapered PCF (NA $\approx 0.38$).
(An additional advantage of the end-tapered fiber is greater mechanical stability.)
This process performed by the fiber vendor is known as end-sealing, and was developed recently to minimise damage to the PCF ends due to high power laser pulses.
The end of the PCF is heated to collapse the air holes surrounding the core.
As a result, the PCF core area expands smoothly to form a cone shape.
The transverse spatial profile of the light exiting from the end-sealed PCF matches well with a Gaussian beam.

To reduce the influence of uncorrelated single photons due to Raman Scattering, the pump wavelength is tuned to $741.7 \pm 0.06$ nm, which is slightly blue of the zero-dispersion wavelength \cite{fan07,fulconis07}.
This improves photon-pair purity without much loss in the pair production rate by avoiding the peak of the Raman Scattering.

The pump is a pulsed Ti:Sapphire laser (repetition rate of 76 MHz, and pulse duration of 8 ps).  
Phase matching results in the peak of the signal and idler wavelengths being emitted at 690.4 nm and 801.2 nm respectively. 
After spectral filtering, photons are collected into single-mode fibers and detected by Si avalanche photo-diodes (APDs).
Electronic signals from the detectors are sent to a start-stop data acquisition system for coincidence detection (5 ns time window), with the signal photons providing the start trigger.

The ends of the PCF are not anti-reflection coated resulting in a calculated emission efficiency of $\eta_{fiber}=96\%$.
The collapsed fiber ends, with their resulting larger optical modes, allow the use of simple and inexpensive anti-reflection coated aspheric lenses (NA=0.4, measured transmittance $\eta_{lens}=98\pm1\%$) to couple light into and out of the fiber, rather than high-magnification microscope objectives \cite{goldschmidt08,fan07}.
Note that for microscope objectives to achieve the same level of transmittance requires significantly more expensive custom anti-reflection coatings, so in practice this is often not done.
To efficiently select the photons in the signal and idler channels within a narrow spectral bandwidth, and to reject the bright pump light, reflection Bragg gratings are used (two gratings per channel).
Similar to the design of a fiber Bragg grating, the reflection Bragg grating filter (from OptiGrate \cite{optigrate,disclaimer}) used in our experiment is a Photo-Thermo-Refractive glass plate (which is silicate glass doped with silver, cerium and fluorine) with its refractive index periodically modulated over the entire depth of the plate.
When a light beam incident on this grating has a wavelength resonant with the grating structure, the beam is reflected with a peak wavelength efficiency of $98\%$ \cite{gratingref}.
When operated in free-space mode (as in our experiment), the reflectance wavelength can be tuned slightly ($\approx$ 1\%) by adjusting the incident angle of the light beam onto the filter.

The measured suppression of the pump with a single grating is $\approx 90$ dB, giving a combined suppression of 180 dB from the two gratings in series.
Typical pump power is $\approx 1$ mW; at this level the combined suppression results in negligible pump light at the APDs compared to the photon-pair count rates of several thousands per second.

The full-width-at-half-maximum (FWHM) for each signal filter grating is 0.17 nm and each idler filter is 0.3 nm (taken from the manufacturer's datasheet).
The measured peak reflectance for each grating is $98 \pm 1$\%.
The actual grating profile is box-like as it falls off faster than a Gaussian, and the peak of the profile is much broader compared to a Gaussian.
Since the photon-pairs are carved from a broad SFWM spectrum, we approximate the gratings' box-like profiles by using the transmission of a Gaussian band-pass averaged over their nominal FWHM.
The average grating efficiency (for its box-like profile), $\eta_{grating}$, is estimated to be 80\%.
Using two gratings in reflectance leads to a combined efficiency of $\eta_{grating}^2 =64\%$ over that FWHM.
Light transmitted through the grating suffers $\approx 10$\% scattering loss.

However, there is an additional loss mechanism due to the finite bandwidth of the pump being comparable to the filter bandwidths.
From the combination of pump and filter bandwidths, we calculate the signal photons to have a spectral width of $\approx 0.39$ nm.
Similarly the bandwidth for the idler photons is $\approx 0.45$ nm.
Using the nominal filter bandwidth, the overall spectral selection efficiency, $\eta_{spectral}$, for the signal photons is $\frac{0.17}{0.39}\times\eta_{grating}^2 \approx 28\%$, while the efficiency for the idler arm is $0.9\times\frac{0.3}{0.45} \times \eta_{grating}^2\approx 38\%$ (which suffer the additional scattering loss due to having to traverse the signal grating).

The single spatial-mode is defined by the single-mode collection fibers.  
Because of a lens common to the signal and idler paths, the signal and idler coupling efficiencies, $\eta_{coupling}$, could not be optimized independently.
Given this tradeoff we matched the signal and idler efficiencies at $50\pm 1\%$.  This efficiency could increase with the use of mode-matching optics such as cylindrical lenses or more sophisticated adaptive optics techniques that have demonstrated 97\% coupling efficiency between single-mode fibers \cite{scott08},
although higher order azimuthal assymmetries may result in a small but fundamental limit to achieving 100\% coupling efficiencies.
Together, these values determine the single photon extraction efficiency.  
For signal photons, this is $\eta_{signal}=\eta_{coupling}\eta_{spectral} \eta_{lens} \eta_{fiber}= 13\pm 0.6\%$ , whereas for idler photons $\eta_{idler} = \eta_{coupling}\eta_{spectral} \eta_{lens}\eta_{fiber}=18\pm 0.6\%$.
The overall photon-pair extraction efficiency is $\eta_{pair}=\eta_{signal} \eta_{idler} \approx 2.3\%$, which is higher than a previous implementation \cite{goldschmidt08}, inferred to be 0.7\% .

The single photon detectors have measured detection efficiencies, $\eta_{det}$, of $56\pm0.6$\% at 690.4 nm and $43\pm0.4\%$ at 801.2 nm.  Thus, taking into account the above values for $\eta_{det}$, the single photon detection efficiency for signal photons is $7.3\pm0.3\%$ and for idler photons is $7.7\pm0.3\%$.  This is close to the maximum observed coincidence-to-singles ratio of $9.57$ with a statistical uncertainty of $\pm 0.02\%$.
This higher value is likely due to our conservative modeling of the bandwidth efficiency.

Putting all the efficiencies together, we find that with our single-mode pair source, we detect $\approx$ 0.6\% of all photon-pairs generated in the selected spectral band.  This is higher than the collection efficiencies of a previous implementation that achieved an efficiency of 0.21\% \cite{goldschmidt08} (Table 1).
Compared to source \cite{fulconis05} in Table 1, we see that source achieving a higher extraction efficiency due to broader bandwidth filters, but we note two important points.
By moving to broader bandwidth Bragg gratings, we would expect the extraction efficiencies to become comparable to source \cite{fulconis05}, while the terrific pump rejection afforded by the gratings is an important benefit.
Ultimately, a hybrid system consisting of a broader band Bragg grating, for initial suppression of the pump light, followed by a box-like etalon filter might be the optimal solution to maximize extraction efficiency while suppressing unwanted background.
(This was actually the spectral selection technique that we used in the subsequent experiments reported in sections 3 and 4.)

\subsection{Measured purity}
Higher pair detection efficiency together with a narrower selected bandwidth enables a high purity photon-pair source.
The photon-pair purity is important, as it determines the background caused by erroneously identified pairs, thus limiting the number of useful pairs for experiments.
This purity can be estimated using either the coincidence-to-accidentals ratio (C/A), or from the second-order correlation function $g^{(2)}(0)$ \cite{loudonbook}.
C/A is commonly used \cite{chen06,lee06} because the accidentals rate is a direct measure of the background level.
The C/A value can be obtained directly from two-fold coincidence measurement setups such as the one depicted in Fig. \ref{fig:fiberscheme}.

\begin{figure} \centerline{\scalebox{0.3} { \includegraphics{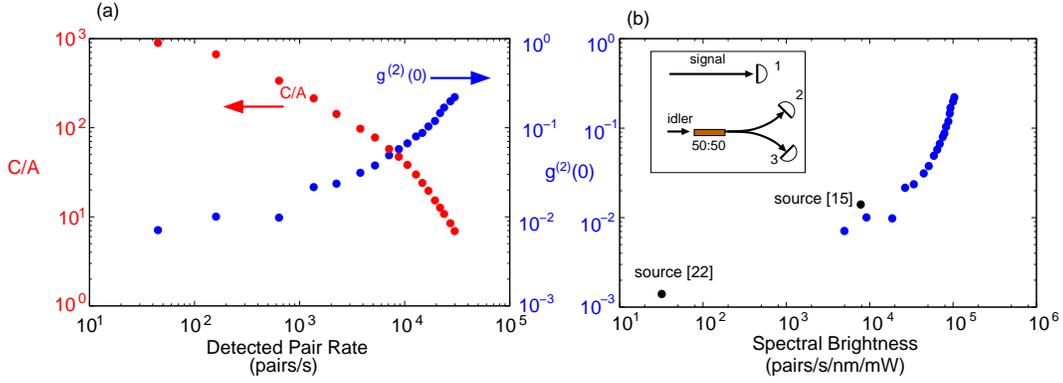}}} \caption{\label{fig:puritycompare} (Color online) Two measures of photon-pair purity: the coincidence-to-accidentals ratio (C/A) and $g^{(2)}(0)$.  (a) Photon-pair purity dependence on the detected pair rate.  (b) $g^{(2)}(0)$ versus {\it raw} spectral brightness.  The inset indicates the detection arrangement for obtaining $g^{(2)}(0)$.  The signal photon acts as a herald, while the idler photons are sent into a polarization neutral 50:50 beamsplitter.  The rate of three-fold and two-fold coincidences determine the value of $g^{(2)}(0)$.  In both (a) and (b) the x-axis was obtained by varying pump power, with higher power yielding higher pair rates and higher spectral brightness.  } \end{figure} 

\begin{table}[b] \caption{Selected data points from Fig. \ref{fig:puritycompare}(b) for comparing $g^{(2)}(0)$ values between different sources.  Increasing the pump repetition rate, but keeping peak pulse power constant, it should be possible to increase the pair production rate while maintaining the level of $g^{(2)}(0)$.} \hspace{-0cm} \begin{tabular}{p{1.7cm}x{2.3cm}x{2cm}x{1.6cm}x{3.4cm}} \hline\hline & \multirow{2}{*}{$g^{(2)}(0)$} & Detected Rate & Bandwidth & Spectral Brightness\tabularnewline & & (pairs $\mbox{s }^{-1}$) & (nm) & (pairs $\mbox{s}^{-1} \mbox{nm}^{-1} \mbox{mW}^{-1}$)\tabularnewline\hline \multirow{2}{*}{Our Source} & $0.007\pm 0.005$ & 45 & 0.17 & 5,300\tabularnewline & $0.03\pm 0.001$ & 3,800 & 0.17 & 44,700\tabularnewline \hline Source \cite{goldschmidt08} & $0.014\pm0.002$ & 350 & 0.9 & 7,800\tabularnewline \hline Source \cite{fasel04} & $0.0014\pm0.0003$ & 350 & 6.9 & 32\tabularnewline \hline\hline \end{tabular} \end{table}
To determine C/A, it is necessary to estimate the rate of accidental coincidences.
To a first approximation, the rate of coincidences between photons from different pump pulses is a good estimate of the accidentals rate.
Our start-stop acquisition system lets us monitor coincidence and accidental counts simultaneously.
The observed C/A taken as the pump power (and detected pair rate) was varied for our PCF is shown in Fig. \ref{fig:puritycompare}(a).  

The second-order correlation function is a direct measure of the presence of multiple photons (per pulse) in the signal and idler channels.
When there is only one pair per pulse, $g^{(2)}(0)=0$; in practice this is never achieved because of detector dark counts.
A semiclassical theory of light establishes a lower limit of 1 for the $g^{(2)}(0)$ of classical coherent light sources \cite{loudonbook}.
Thus we would expect that the $g^{(2)}(0)$ values for our source would lie somewhere between 0 and 1, when operated with low pump power.

In our $g^{(2)}(0)$ measurement scheme (inset of Fig. \ref{fig:puritycompare}(b)),
the idler photons are incident on a 50:50 fiber beamsplitter whose output ports are sent to APDs.  
The detector in the signal arm acts as the herald for a three-fold coincidence.
Following a simple model described in \cite{fasel04}, the correlation function is \begin{equation}g^{(2)}(0) = \frac{4C_{123} C_1}{(C_{12} + C_{13})^2},\end{equation}
where the three-fold coincidence rate is $C_{123}$, the rate of signal photons is $C_1$, and the two-fold rate between signal and idler detectors are $C_{12}$ and $C_{13}$, 
The three-fold coincidences were detected with an electronic circuit based on field programmable gate array (FPGA) technology \cite{fpga}.
The measured correlation values (Fig. \ref{fig:puritycompare}) are much less than 1, signifying the nonclassical nature of the emitted light.

Figure \ref{fig:puritycompare} (a) links photon-pair purity with the detected pair rate.  From the figure, a detection rate of 45 pairs $\mbox{s}^{-1}$ ($\approx 0.05$ mW) has  $g^{(2)}(0)=0.007 \pm 0.005$, and a C/A = 900 (C/A $\rightarrow \infty$ for no background accidentals).  When pump power is increased to 0.5 mW, the detected pair rate is 3,800 pairs $\mbox{s}^{-1}$, $g^{(2)}(0)=0.03 \pm 0.001$ and C/A=100.  For comparison, we consider the lowest noise photon-pair source we found in the literature (based on SPDC \cite{fasel04}).
This source exhibits a $g^{(2)}(0)$ of 0.0014 $\pm 0.0003$ at a rate of 350 pairs $\mbox{s}^{-1}$ and bandwidth of 6.9 nm.
Although reference \cite{fasel04} does not provide the observed pair rate, this value may be inferred from the data that was published in the paper.
To compare pair rates between sources, we normalize to the pump power and collection bandwidth and determine the {\it raw} spectral brightness of our source to be 44,700 pairs $\mbox{(s nm mW)}^{-1}$  at 0.5 mW of pump power (Fig. \ref{fig:puritycompare}(b)).  
Selected points from Fig. \ref{fig:puritycompare}(b) are presented in Table 2.  
\section{Polarization-entangled photon-pairs}

\begin{figure} \centerline{\scalebox{0.39} { \includegraphics{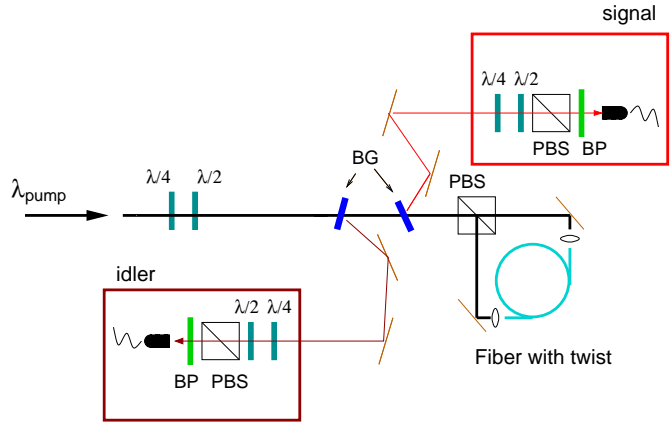}}} \caption{\label{fig:entanglement} (Color online) Schematic of the polarization-entangled photon-pair source based on a $90^o$ twist of the photonic-crystal fiber.  The PCF is pumped in both directions.  A single Bragg grating (BG) selects for each of the signal and idler; to suppress residual pump light highly transmissive ($>99\%$) bandpass (BP) filters are used.  The entangled state is analyzed using a combination of quarter-wave ($\frac{\lambda}{4}$) and half-wave ($\frac{\lambda}{2}$) plates together with a polarizing beam splitter (PBS).  } \end{figure} 

Here, we characterize a polarization-entangled pair source based on an end-tapered PCF.
This is done by using a coherent superposition of \ket{$\mbox{HH}$} and $|\mbox{VV}\rangle$, where $|\mbox{HH}\rangle$ and $|\mbox{VV}\rangle$ represent the horizontal and vertical polarization states of photon-pairs.  
Such a superposition may be generated by placing the PCF in a Sagnac loop as reported in \cite{fiberent07,fulconis07} (Fig. \ref{fig:entanglement}).

To generate the orthogonal polarization states, only one principal axis of the PCF is pumped from both ends.  
The axis at one end was aligned to match with the H output of a polarizing beam splitter (PBS).  
The PCF was twisted so that the axis orientation at the other end is matched with the V output of the PBS.  
The extinction ratio of a PCF-based Sagnac loop is better than 200:1.
Light from the pulsed laser is split by the PBS and pumps the PCF in two directions.
By controlling the polarization of the pump (and hence its splitting ratio at the PBS), it is possible to balance the pair production from both outputs of the PCF.
This is necessary because the photon-pair production rates for the two pump directions are not equal (by $\approx$ 20\%), despite having the same level of inserted pump power (most probably due to non-uniformity in the PCF).  

The recombination of light from both ends at the PBS generates the superposition of $|\mbox{HH}\rangle$ and $|\mbox{VV}\rangle$, producing a polarization-entangled photon-pair state.
Using only a single Bragg grating for each wavelength, we are able to separate the desired signal and idler photons from the rest of the light output.
Sufficient suppression of pump light was achieved with the help of an additional bandpass filter.  
These filters are centered on 800 nm (FWHM $\approx 12$ nm) and 692 nm (FWHM $\approx 40$ nm) respectively, both having a measured transmission efficiency of 99\% and a box-like spectral selection profile.

\begin{figure} \centerline{\scalebox{0.32} { \includegraphics{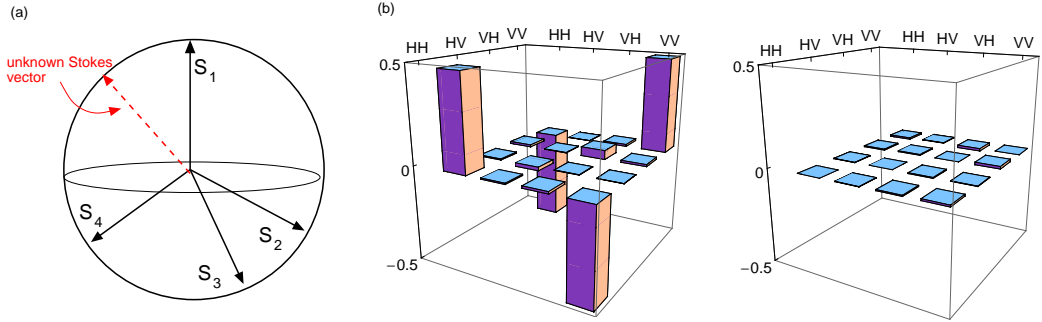}}} \caption{\label{fig:tetrahedroncombo} (Color online) (a) Illustrates the concept of minimal and optimal tomography for an unknown Stokes vector.  (b) A graphical representation of the density matrix obtained using minimal and optimal quantum state tomography.  The real part of the matrix is on the left; the imaginary part is the right.  The magnitude of the components of the imaginary part are less than 0.013.  The fidelity to the $\Phi^-$ Bell state is $97 \pm 1\%$.  } \end{figure}

The polarization-entangled state can be completely characterized by quantum state tomography.
We used a tomographic technique that is known to be minimal and optimal \cite{mqt}.
The key point of this technique is that the polarization state of light is described by a Stokes vector which has only three independent variables \cite{ambi1,ambi2}. 
Such a Stokes vector can be illustrated on a Poincare sphere (Fig. \ref{fig:tetrahedroncombo}(a)).
The direction and magnitude of any Stokes vector is completely determined by its overlap with four reference vectors in the sphere.
In contrast, standard polarimetry requires six overlap measurements \cite{hechttxt}.
It was further shown that when these reference vectors define a tetrahedron in the Poincare sphere, the characterization rate is optimized \cite{mqt}.
Hence, characterization of single photon polarization states requires only 4 projective measurements \cite{jmoling06}.
This is of particular importance when considering N-photon states, where the number of projective measurements grows as $4^N$ \cite{praling06}, in contrast to standard polarimetry that grows as $6^N$.

To characterize our photon-pair state, we needed to monitor 16 separate two-fold coincidences.
This was done by projecting the signal photons sequentially onto the 4 reference polarizations states; the reference states were prepared by rotating the half-wave and quarter-wave plates (Fig. \ref{fig:entanglement}) to the angles described in \cite{ambi2}.
For each projection state of the signal, the idler photons are also projected onto the same four reference states.
In this way, we obtain the 16 combinations of coincidences that are sufficient to obtain a photon-pair Stokes vector that can be converted into a density matrix \cite{praling06}.

The source is typically operated at room temperature with a pump power of 1 mW (before the PBS), and the detected pair rate is $\approx 2,800 \mbox{ s}^{-1}$.
The detected rate is lower than with a single pump direction setup because the spatial mode shapes from the two fiber ends are slightly different, causing additional loss in coupling into single-mode fibers that define the useful spatial mode as well as transmit light to the APDs.
The density matrix of our photon-pair state (at 1 mW of pump power and {\it without} correcting for accidentals) is reconstructed and shown in Fig. \ref{fig:tetrahedroncombo}(b).  
The fidelity of this density matrix to the maximally entangled Bell state, $\Phi^-=\frac{1}{\sqrt{2}}\left (|\mbox{HH}\rangle - |\mbox{VV}\rangle \right )$, is $97\pm 1$\% (error propagation assumes a Poissonian noise model and standard error is used).
The tangle is one method of quantifying the degree of entanglement, and from our measured density matrix the tangle is found to be $92 \pm 2$\%.

\section{Heralded indistinguishable single photons}
\begin{figure} \centerline{\scalebox{0.3} { \includegraphics{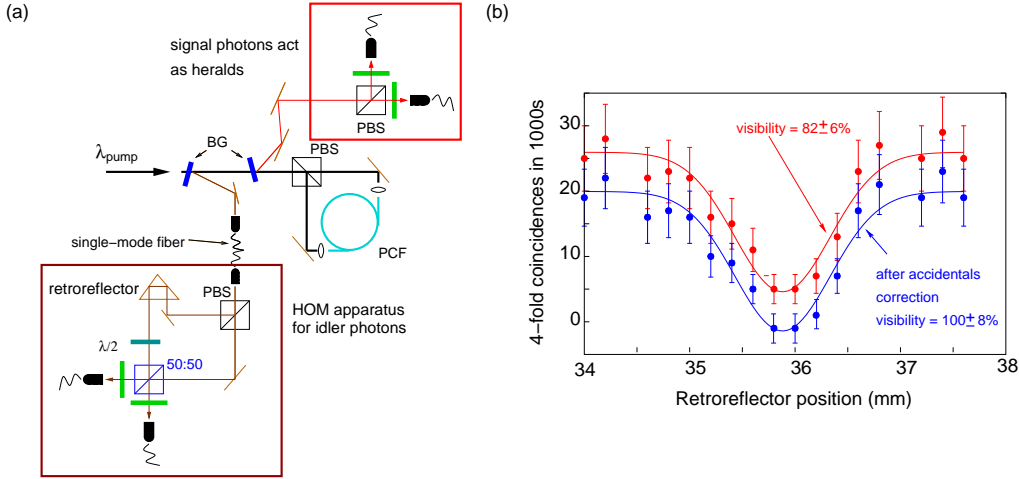}}} \caption{\label{fig:homcombo}(Color online) (a) Scheme for measurement of the Hong-Ou-Mandel Interference dip.  The signal photons act as heralds for the idler photons.  For interference to take place, the idler photons are set to H polarization. (b)  The observed HOM dip at $\approx$ 1 mW of pump power in each arm of the PCF.  When corrected for accidentals, the HOM dip is compatible with unit visibility.  } \end{figure} 

Another possible use of the end-tapered PCF is as a source of heralded indistinguishable single photons.
We note that a single PCF that is pumped bi-directionally acts as two sources of heralded single photons (similar to some SPDC experiments where a nonlinear crystal is pumped in two directions \cite{pan98}).
When combined with the spectral filtering described in previous sections, the idler (and signal) photons generated from either end of a PCF are effectively indistinguishable.
We demonstrate this by performing the classic Hong-Ou-Mandel (HOM) photon interference measurement \cite{hong87}.

This experiment (Fig. \ref{fig:homcombo}(a)) is a modification of the setup used to generate the polarization-entangled pairs where two pairs of photons are collected, one pair from each output of the PCF. 
Using the FPGA-based counting system, the overall rate of four-fold coincidences was monitored.
To act as heralds for their respective idler partner, the signal photons were split off via a PBS.
The idler photons were also sent through a PBS to identify their polarization.
Idler photons leaving from the V-polarized port of the Sagnac loop were rotated with a half-wave plate to match the polarization of the photons leaving from the H port.
The photons are then interfered on a 50:50 beamsplitter.
For a well defined spatial mode and maximal spatial overlap of idler photons at the 50:50 beamsplitter, we used a single-mode fiber between the grating and the first PBS in the Hong-Ou-Mandel setup.
The degree of temporal overlap between the idler wavepackets was controlled by moving a retroreflector to adjust the path delay.  
The rate of four-fold coincidences is recorded against the position of this retroreflector.

The main source of noise in this measurement is multiple pair generation from a single pump pulse direction causing a background that reduces the visibility of the HOM interference.  
The background rate can be determined by sequentially blocking each input port of the 50:50 beamsplitter, and adding up the remaining four-fold coincidences, and is found to be $\approx 0.005 \mbox{ s}^{-1}$.
At 1 mW of pump power ($g^{(2)}(0)\approx 0.08$ and C/A $\approx 30$), the average rate of four-fold coincidences was $\approx 0.03 \mbox{ s}^{-1}$.
These should be compared against the state-of-the art in four-fold coincidence generation via SPDC ($0.28 \mbox{ s}^{-1}$ \cite{mosley08}) or SFWM ($\approx 2.5 \mbox{ s}^{-1}$ \cite{cohen08}).
It would be very interesting to combine the SFWM design techniques in \cite{cohen08} with end-tapered PCFs to obtain even brighter sources of four-fold coincidences.

When the idler photon wavepackets had maximal spatial overlap, a dip was obtained in the raw four-fold coincidence rate with a visibility of $82 \pm 6$\%.
When the estimated four-fold accidentals rate is subtracted from the observed dip, we find that the dip approaches unit visibility  (Fig. \ref{fig:homcombo}(b)).
This result indicates that by operating at lower pump power (but requiring longer integration times or higher repetition rates), it is possible to herald purely indistinguishable single photons.

\section{Conclusion}
We have reported a substantial improvement in the purity ($g^2(0)$ and entanglement fidelity) of single-mode photon-pairs made using PCFs.
This improvement was brought about by the use of end-tapered PCFs and novel spectral filters.
We studied the photon-pair emission characteristics of a pulse-pumped end-tapered PCF, and presented the dependence of count rates and noise on pump power.

The demonstrated combination of high count rate and low background noise shows that it is possible to build bright sources of high quality polarization-entangled photon-pairs by using an end-tapered PCF in a Sagnac loop, along with Bragg gratings for spectral selection and pump rejection.
Our source displays higher photon-pair purity, indistinguishability and polarization-entanglement quality compared to previous implementations of fiber-based photon-pair sources \cite{fiberent07,goldschmidt08,fulconis07}. 
Furthermore, the increased mode area of end-tapered fibers improves the mechanical stability and coupling efficiency of the bi-directionally pumped fibers.
In addition the use of broader bandwidth Bragg gratings, more appropriate for the pump bandwidth used, would allow spectral efficiencies comparable to an etalon filtered source.
Ultimately, a hybrid system consisting of one high-efficiency notch filter followed by a single broader band Bragg grating might be the optimal solution to maximize extraction efficiency while suppressing unwanted background.

One further direction is the use of PCFs \cite{garay07,cohen08,halder09,soller09} to generate factorizable states, which in principle, can remove the need for spectral filtering.
In reference \cite{halder09}, a custom PCF was used to generate nearly factorizable states, allowing a four-fold rate of 0.3 $\mbox{s}^{-1}$ to be achieved.
The combination of factorizable states, end-tapers and mode-matching optics could lead to sources with even higher levels of detected pair brightness.
Furthermore, PCFs in such arrangements need not be limited to photon-counting experiments since it is also possible to use end-tapered fibers to generate squeezed light \cite{heersink05,milanovic09}.
Together, these possibilities highlight the potential of end-tapered PCFs as a robust source of non-classical light.

\section*{Acknowledgments}
This work has been supported in part by the Intelligence Advanced Research Projects Activity (IARPA) entangled photon source program.

\begin{thebibliography}{36}
\expandafter\ifx\csname natexlab\endcsname\relax\def\natexlab#1{#1}\fi
\expandafter\ifx\csname bibnamefont\endcsname\relax
  \def\bibnamefont#1{#1}\fi
\expandafter\ifx\csname bibfnamefont\endcsname\relax
  \def\bibfnamefont#1{#1}\fi
\expandafter\ifx\csname citenamefont\endcsname\relax
  \def\citenamefont#1{#1}\fi
\expandafter\ifx\csname url\endcsname\relax
  \def\url#1{\texttt{#1}}\fi
\expandafter\ifx\csname urlprefix\endcsname\relax\def\urlprefix{URL }\fi
\providecommand{\bibinfo}[2]{#2}
\providecommand{\eprint}[2][]{\url{#2}}

\bibitem{coinc70}
\bibinfo{author}{\bibfnamefont{D.~C.} \bibnamefont{Burnham}} \bibnamefont{and}
  \bibinfo{author}{\bibfnamefont{D.~L.} \bibnamefont{Weinberg}},\bibinfo{title}{ ``Observation of Simultaneity in Parametric Production of Optical Photon Pairs,"}
  \bibinfo{journal}{Phys. Rev. Lett.} \textbf{\bibinfo{volume}{25}},
  \bibinfo{pages}{84-87} (\bibinfo{year}{1970})
  .

\bibitem{kurtsiefer01}
\bibinfo{author}{\bibfnamefont{C.}~\bibnamefont{Kurtsiefer}},
  \bibinfo{author}{\bibfnamefont{M.}~\bibnamefont{Oberparleiter}},
  \bibnamefont{and}
  \bibinfo{author}{\bibfnamefont{H.}~\bibnamefont{Weinfurter}},
\bibinfo{title}{``High-efficiency entangled photon pair collection in type-II parametric fluorescence,"}
  \bibinfo{journal}{Phys. Rev. A }
  \textbf{\bibinfo{volume}{64}}, \bibinfo{pages}{023802}
  (\bibinfo{year}{2001}).

\bibitem{fed07}
\bibinfo{author}{\bibfnamefont{A.}~\bibnamefont{Fedrizzi}},
  \bibinfo{author}{\bibfnamefont{T.}~\bibnamefont{Herbst}},
  \bibinfo{author}{\bibfnamefont{A.}~\bibnamefont{Poppe}},
  \bibinfo{author}{\bibfnamefont{T.}~\bibnamefont{Jennewein}},
  \bibnamefont{and}
  \bibinfo{author}{\bibfnamefont{A.}~\bibnamefont{Zeilinger}},
\bibinfo{title}{``A wavelength-tunable fiber-coupled source of narrowband entangled photons,"}
  \bibinfo{journal}{Opt. Express} \textbf{\bibinfo{volume}{15}},
  \bibinfo{pages}{15377-15386} (\bibinfo{year}{2007}).

\bibitem{uren04}
\bibinfo{author}{\bibfnamefont{A.~B.} \bibnamefont{U'Ren}},
  \bibinfo{author}{\bibfnamefont{C.}~\bibnamefont{Silberhorn}},
  \bibinfo{author}{\bibfnamefont{K.}~\bibnamefont{Banaszek}}, \bibnamefont{and}
  \bibinfo{author}{\bibfnamefont{I.~A.} \bibnamefont{Walmsley}},
\bibinfo{title}{``Efficient Conditional Preparation of High-Fidelity Single Photon States for Fiber-Optic Quantum Networks,"}
  \bibinfo{journal}{Phys. Rev. Lett.} \textbf{\bibinfo{volume}{93}},
  \bibinfo{eid}{093601} (\bibinfo{year}{2004}).

\bibitem{fiorentino07}
\bibinfo{author}{\bibfnamefont{M.}~\bibnamefont{Fiorentino}},
  \bibinfo{author}{\bibfnamefont{S.~M.} \bibnamefont{Spillane}},
  \bibinfo{author}{\bibfnamefont{R.~G.} \bibnamefont{Beausoleil}},
  \bibinfo{author}{\bibfnamefont{T.~D.} \bibnamefont{Roberts}},
  \bibinfo{author}{\bibfnamefont{P.}~\bibnamefont{Battle}}, \bibnamefont{and}
  \bibinfo{author}{\bibfnamefont{M.~W.} \bibnamefont{Munro}},
\bibinfo{title}{``Spontaneous parametric down-conversion in periodically poled KTP waveguides and bulk crystals,"}
  \bibinfo{journal}{Opt. Express} 
\textbf{\bibinfo{volume}{15}}, \bibinfo{pages}{7479-7488} (\bibinfo{year}{2007}).

\bibitem{chen09}
\bibinfo{author}{\bibfnamefont{J.}~\bibnamefont{Chen}},
  \bibinfo{author}{\bibfnamefont{A.~J.} \bibnamefont{Pearlman}},
  \bibinfo{author}{\bibfnamefont{A.}~\bibnamefont{Ling}},
  \bibinfo{author}{\bibfnamefont{J.}~\bibnamefont{Fan}}, \bibnamefont{and}
  \bibinfo{author}{\bibfnamefont{A.}~\bibnamefont{Migdall}},
\bibinfo{title}{``A versatile waveguide source of photon pairs for chip-scale quantum information processing,"}
  \bibinfo{journal}{Opt. Express} 
\textbf{\bibinfo{volume}{17}}, \bibinfo{pages}{6727-6740} (\bibinfo{year}{2009}).

\bibitem{christ09}
\bibinfo{author}{\bibfnamefont{A.}~\bibnamefont{Christ}},
  \bibinfo{author}{\bibfnamefont{K.}~\bibnamefont{Laiho}},
  \bibinfo{author}{\bibfnamefont{A.}~\bibnamefont{Eckstein}},
  \bibinfo{author}{\bibfnamefont{T.}~\bibnamefont{Lauckner}},
  \bibinfo{author}{\bibfnamefont{P.~J.} \bibnamefont{Mosley}},
  \bibnamefont{and}
  \bibinfo{author}{\bibfnamefont{C.}~\bibnamefont{Silberhorn}},
\bibinfo{title}{``Spatial modes in waveguided parametric downconversion,"}
  \bibinfo{journal}{arXiv:0904.4668v1}  (\bibinfo{year}{2009}).

\bibitem{karpinski09} \bibinfo{author}{\bibfnamefont{M.}~\bibnamefont{Karpinski}}, \bibinfo{author}{\bibfnamefont{C.}~\bibnamefont{Radzewicz}}, \bibnamefont{and} \bibinfo{author}{\bibfnamefont{K.}~\bibnamefont{Banaszek}}, \bibinfo{title}{``Experimental characterization of three-wave mixing in a multimode nonlinear KTiOPO\_4 waveguide,"}  \bibinfo{journal}{Appl. Phys. Lett.} \textbf{\bibinfo{volume}{94}} \bibinfo{pages}{181105} (\bibinfo{year}{2009}).

\bibitem{fiorentino02} \bibinfo{author}{\bibfnamefont{M.}~\bibnamefont{Fiorentino}}, \bibinfo{author}{\bibfnamefont{P.}~\bibfnamefont{L.}~\bibnamefont{Voss}}, \bibinfo{author}{\bibfnamefont{J.}~\bibfnamefont{E.}~\bibnamefont{Sharping}}, \bibnamefont{and} \bibinfo{author}{\bibfnamefont{P.}~\bibnamefont{Kumar}}, \bibinfo{title}{``All-Fiber Photon-Pair Source for Quantum Communications,"}  \bibinfo{journal}{IEEE Phot. Tech. Lett.} \bibinfo{pages}{983-985}  (\bibinfo{year}{2002}).

\bibitem{rarity05} \bibinfo{author}{\bibfnamefont{J.}~\bibfnamefont{G.}~\bibnamefont{Rarity}}, \bibinfo{author}{\bibfnamefont{J.}~\bibnamefont{Fulconis}}, \bibinfo{author}{\bibfnamefont{J.}~\bibnamefont{Duligall}}, \bibinfo{author}{\bibfnamefont{W.}~\bibfnamefont{J.}~\bibnamefont{Wadsworth}}, \bibnamefont{and} \bibinfo{author}{\bibfnamefont{P.}~\bibfnamefont{St.}~\bibfnamefont{J.}\bibnamefont{Russell}}, \bibinfo{title}{``Photonic crystal fiber source of correlated photon pairs,"}  \bibinfo{journal}{Opt. Exp.} \textbf{\bibinfo{volume}{13}},
\bibinfo{pages}{534-544} (\bibinfo{year}{2005}).

\bibitem{dyer09} \bibinfo{author}{\bibfnamefont{S.}~\bibfnamefont{D.}~\bibnamefont{Dyer}}, \bibinfo{author}{\bibfnamefont{B.}~\bibnamefont{Baek}}, \bibnamefont{and} \bibinfo{author}{\bibfnamefont{S.}~\bibfnamefont{W.}~\bibnamefont{Nam}}, \bibinfo{title}{``High-brightness, low-noise, all-fiber photon pair source,"}  \bibinfo{journal}{Opt. Exp.}  \textbf{\bibinfo{volume}{17}} \bibinfo{pages}{10290-10297}(\bibinfo{year}{2009}).

\bibitem{fan05}
\bibinfo{author}{\bibfnamefont{J.}~\bibnamefont{Fan}},
  \bibinfo{author}{\bibfnamefont{A.}~\bibnamefont{Migdall}}, \bibnamefont{and}
  \bibinfo{author}{\bibfnamefont{L.~J.} \bibnamefont{Wang}},
\bibinfo{title}{``Efficient generation of correlated photon pairs in a microstructure fiber,"}
  \bibinfo{journal}{Opt. Lett.} \textbf{\bibinfo{volume}{30}},
  \bibinfo{pages}{3368-3370} (\bibinfo{year}{2005}).

\bibitem{takesue05}
\bibinfo{author}{\bibfnamefont{H.}~\bibnamefont{Takesue}} \bibnamefont{and}
  \bibinfo{author}{\bibfnamefont{K.}~\bibnamefont{Inoue}},
\bibinfo{title}{``1.5-$\mu$m band quantum-correlated photon pair generation in dispersion-shifted fiber: suppression of noise photons by cooling fiber,"}
  \bibinfo{journal}{Opt. Express} \textbf{\bibinfo{volume}{13}},
  \bibinfo{pages}{7832-7839} (\bibinfo{year}{2005}).

\bibitem{chen06}
\bibinfo{author}{\bibfnamefont{J.}~\bibnamefont{Chen}},
  \bibinfo{author}{\bibfnamefont{K.~F.} \bibnamefont{Lee}},
  \bibinfo{author}{\bibfnamefont{C.}~\bibnamefont{Liang}}, \bibnamefont{and}
  \bibinfo{author}{\bibfnamefont{P.}~\bibnamefont{Kumar}},
\bibinfo{title}{``Fiber-based telecom-band degenerate-frequency source of entangled photon pairs,"}
  \bibinfo{journal}{Opt. Lett.} \textbf{\bibinfo{volume}{31}},
  \bibinfo{pages}{2798-2800} (\bibinfo{year}{2006}).

\bibitem{lee06}
\bibinfo{author}{\bibfnamefont{K.~F.} \bibnamefont{Lee}},
  \bibinfo{author}{\bibfnamefont{J.}~\bibnamefont{Chen}},
  \bibinfo{author}{\bibfnamefont{C.}~\bibnamefont{Liang}},
  \bibinfo{author}{\bibfnamefont{X.}~\bibnamefont{Li}},
  \bibinfo{author}{\bibfnamefont{P.~L.} \bibnamefont{Voss}}, \bibnamefont{and}
  \bibinfo{author}{\bibfnamefont{P.}~\bibnamefont{Kumar}},
\bibinfo{title}{``Generation of high-purity telecom-band entangled photon pairs in dispersion-shifted fiber,"}
  \bibinfo{journal}{Opt. Lett.} \textbf{\bibinfo{volume}{31}},
  \bibinfo{pages}{1905-1907} (\bibinfo{year}{2006}).

\bibitem{russell03}
\bibinfo{author}{\bibfnamefont{P.}~\bibnamefont{Russell}},
\bibinfo{title}{``Photonic Crystal Fibers,"}
  \bibinfo{journal}{Science} \textbf{\bibinfo{volume}{299}},
  \bibinfo{pages}{358-362} (\bibinfo{year}{2003}).

\bibitem{fiberent07}
\bibinfo{author}{\bibfnamefont{J.}~\bibnamefont{Fan}},
  \bibinfo{author}{\bibfnamefont{M.~D.} \bibnamefont{Eisaman}},
  \bibnamefont{and} \bibinfo{author}{\bibfnamefont{A.}~\bibnamefont{Migdall}},
\bibinfo{title}{``Bright phase-stable broadband fiber-based source of polarization-entangled photon pairs,"}  \bibinfo{journal}{Phy. Rev. A }
  \textbf{\bibinfo{volume}{76}}, \bibinfo{pages}{043836}
  (\bibinfo{year}{2007}).

\bibitem{goldschmidt08} \bibinfo{author}{\bibfnamefont{E.~A.} \bibnamefont{Goldschmidt}}, \bibinfo{author}{\bibfnamefont{M.~D.} \bibnamefont{Eisaman}}, \bibinfo{author}{\bibfnamefont{J.}~\bibnamefont{Fan}}, \bibinfo{author}{\bibfnamefont{S.~V.} \bibnamefont{Polyakov}}, \bibnamefont{and} \bibinfo{author}{\bibfnamefont{A.}~\bibnamefont{Migdall}}, \bibinfo{title}{``Spectrally bright and broad fiber-based heralded single-photon source,"}
\bibinfo{journal}{Phys. Rev. A } \textbf{\bibinfo{volume}{78}}, \bibinfo{eid}{013844} (\bibinfo{year}{2008}).

\bibitem{fulconis07}
\bibinfo{author}{\bibfnamefont{J.}~\bibnamefont{Fulconis}},
  \bibinfo{author}{\bibfnamefont{O.}~\bibnamefont{Alibart}},
  \bibinfo{author}{\bibfnamefont{J.~L.} \bibnamefont{O'Brien}},
  \bibinfo{author}{\bibfnamefont{W.~J.} \bibnamefont{Wadsworth}},
  \bibnamefont{and} \bibinfo{author}{\bibfnamefont{J.~G.}
  \bibnamefont{Rarity}},
\bibinfo{title}{``Nonclassical interference and Entanglement Generation Using a Photonic Crystal Fiber Pair Photon Source,"} \bibinfo{journal}{Phys. Rev. Lett.},
  \textbf{\bibinfo{volume}{99}}, \bibinfo{eid}{120501}
  (\bibinfo{year}{2007}).


\bibitem{fulconis05} \bibinfo{author}{\bibfnamefont{J.}~\bibnamefont{Fulconis}}, 
\bibinfo{author}{\bibfnamefont{O.}~\bibnamefont{Alibart}}, 
\bibinfo{author}{\bibfnamefont{W.~J.}~\bibnamefont{Wadsworth}}, 
\bibinfo{author}{\bibfnamefont{P.~St.~J.}~\bibnamefont{Russell}}, 
\bibnamefont{and} \bibinfo{author}{\bibfnamefont{J.~G.}~\bibnamefont{Rarity}},
\bibinfo{title}{``High brightness single mode source of correlated photon pairs using a photonic crystal fiber,"} \bibinfo{journal}{Opt. Express} \textbf{\bibinfo{volume}{13}}, \bibinfo{pages}{7572-7582} (\bibinfo{year}{2005}).

\bibitem{chen05} \bibinfo{author}{\bibfnamefont{J.}~\bibnamefont{Chen}}, \bibinfo{author}{\bibfnamefont{X.}~\bibnamefont{Li}}, \bibnamefont{and} \bibinfo{author}{\bibfnamefont{P.}~\bibnamefont{Kumar}}, 
\bibinfo{title}{``Two-photon-state generation via four-wave mixing in optical fibers,"}
\bibinfo{journal}{Phys. Rev. A} \textbf{\bibinfo{volume}{72}}, \bibinfo{pages}{033801} (\bibinfo{year}{2005}).

\bibitem{fan07}
\bibinfo{author}{\bibfnamefont{J.}~\bibnamefont{Fan}} \bibnamefont{and}
  \bibinfo{author}{\bibfnamefont{A.}~\bibnamefont{Migdall}},
\bibinfo{title}{``A broadband high spectral brightness fiber-based two-photon source,"}
  \bibinfo{journal}{Opt. Express} \textbf{\bibinfo{volume}{15}},
  \bibinfo{pages}{2915-2920} (\bibinfo{year}{2007}).

\bibitem{crystalfibre}
\urlprefix\url{http:\\www.nktphotonics.com},
\bibinfo{title}{Crystal Fibre A/S has been merged into NKT Photonics}.

\bibitem{disclaimer}
\bibinfo{title}{Certain trade names and company products are mentioned in the text or identified in an illustration in order to specify adequately the experimental procedure and equipment used.  In no case does such identification imply recommendation or endorsement by the National Institute of Standards and Technology, nor does it necessarily imply that the products are the best available for the purpose}.

\bibitem{optigrate}
\urlprefix\url{http:\\www.optigrate.com}.

\bibitem{gratingref}
\bibinfo{author}{\bibfnamefont{I.}~\bibnamefont{Ciapurin}},
\bibinfo{author}{\bibfnamefont{L.}~\bibnamefont{Glebov}},
\bibinfo{author}{\bibfnamefont{V.}~\bibnamefont{Smirnov}},
\bibinfo{title}{``Practical Holography XIX:Materials and Applications. Eds: T. H. Jeong, H. Bjelkhagen",}
\bibinfo{journal}{Proceedings of SPIE 5742},
\bibinfo{pages}{183-194},
(\bibinfo{year}{2005}).

\bibitem{scott08}
\bibinfo{author}{\bibfnamefont{S.}~\bibnamefont{Jobling}},
  \bibinfo{author}{\bibfnamefont{K.~T.} \bibnamefont{McCusker}},
  \bibnamefont{and} \bibinfo{author}{\bibfnamefont{P.~G.} \bibnamefont{Kwiat}},
\bibinfo{title}{``Adaptive Optics for Improved Mode-Coupling Efficiencies,"}
  in \emph{\bibinfo{booktitle}{Poster JWA32, Frontiers in Optics}}
  (\bibinfo{year}{2008}).

\bibitem{loudonbook}
\bibinfo{author}{\bibfnamefont{R.}~\bibnamefont{Loudon}},
  \emph{\bibinfo{title}{The Quantum Theory of Light}} (\bibinfo{publisher}{New
  York:Oxford University Press}, \bibinfo{year}{1983}).

\bibitem{fasel04}
\bibinfo{author}{\bibfnamefont{S.}~\bibnamefont{Fasel}},
  \bibinfo{author}{\bibfnamefont{O.}~\bibnamefont{Alibart}},
  \bibinfo{author}{\bibfnamefont{S.}~\bibnamefont{Tanzilli}},
  \bibinfo{author}{\bibfnamefont{P.}~\bibnamefont{Baldi}},
  \bibinfo{author}{\bibfnamefont{A.}~\bibnamefont{Beveratos}},
  \bibinfo{author}{\bibfnamefont{N.}~\bibnamefont{Gisin}}, \bibnamefont{and}
  \bibinfo{author}{\bibfnamefont{H.}~\bibnamefont{Zbinden}},
 \bibinfo{title}{``High-quality asynchronous heralded single-photon source at telecom wavelength,"}
  \bibinfo{journal}{New Journal of Physics} \textbf{\bibinfo{volume}{6}},
  \bibinfo{pages}{163} (\bibinfo{year}{2004}).

\bibitem{fpga}
\urlprefix\url{http://www.nist.gov/fpga}.

\bibitem{mqt}
\bibinfo{author}{\bibfnamefont{J.}~\bibnamefont{Reh\'{a}\v{c}ek}},
  \bibinfo{author}{\bibfnamefont{B.-G.} \bibnamefont{Englert}},
  \bibnamefont{and}
  \bibinfo{author}{\bibfnamefont{D.}~\bibnamefont{Kaszlikowski}},
\bibinfo{title}{``Minimal qubit tomography,"}
  \bibinfo{journal}{Phys. Rev. A }
  \textbf{\bibinfo{volume}{70}}, \bibinfo{eid}{052321}
   (\bibinfo{year}{2004}).

\bibitem{ambi1}
\bibinfo{author}{\bibfnamefont{A.}~\bibnamefont{Ambirajan}} \bibnamefont{and}
  \bibinfo{author}{\bibfnamefont{D.~C.} \bibnamefont{Look}},
\bibinfo{title}{``Optimum angles for a polarimeter: part I,"}
  \bibinfo{journal}{Opt. Eng.} \textbf{\bibinfo{volume}{34}},
  \bibinfo{pages}{(6), 1651} (\bibinfo{year}{1995}).

\bibitem{ambi2}
\bibinfo{author}{\bibfnamefont{A.}~\bibnamefont{Ambirajan}} \bibnamefont{and}
  \bibinfo{author}{\bibfnamefont{D.~C.} \bibnamefont{Look}},
\bibinfo{title}{``Optimum angles for a polarimeter: part II,"}
  \bibinfo{journal}{Opt. Eng.} \textbf{\bibinfo{volume}{34}},
  \bibinfo{pages}{(6), 1656} (\bibinfo{year}{1995}).

\bibitem{hechttxt}
\bibinfo{author}{\bibfnamefont{E.}~\bibnamefont{Hecht}},
  \emph{\bibinfo{title}{Optics, 4th Edition}} (\bibinfo{publisher}{Addison
  Wesley}, \bibinfo{year}{2002}).

\bibitem{jmoling06}
\bibinfo{author}{\bibfnamefont{A.}~\bibnamefont{Ling}},
  \bibinfo{author}{\bibfnamefont{K.~P.} \bibnamefont{Soh}},
  \bibinfo{author}{\bibfnamefont{A.}~\bibnamefont{Lamas-Linares}},
  \bibnamefont{and}
  \bibinfo{author}{\bibfnamefont{C.}~\bibnamefont{Kurtsiefer}},
\bibinfo{title}{``An optimal photon counting polarimeter,"}
  \bibinfo{journal}{Journal of Modern Optics} \textbf{\bibinfo{volume}{56}},
  \bibinfo{pages}{1523-1528} (\bibinfo{year}{2006}).

\bibitem{praling06}
\bibinfo{author}{\bibfnamefont{A.}~\bibnamefont{Ling}},
  \bibinfo{author}{\bibfnamefont{K.~P.} \bibnamefont{Soh}},
  \bibinfo{author}{\bibfnamefont{A.}~\bibnamefont{Lamas-Linares}},
  \bibnamefont{and}
  \bibinfo{author}{\bibfnamefont{C.}~\bibnamefont{Kurtsiefer}},
\bibinfo{title}{``Experimental polarization state tomography using optimal polarimeters,"}
  \bibinfo{journal}{Phys. Rev. A }
  \textbf{\bibinfo{volume}{74}}, \bibinfo{eid}{022309}
   (\bibinfo{year}{2006}).

\bibitem{pan98}
\bibinfo{author}{\bibfnamefont{J.-W.} \bibnamefont{Pan}},
  \bibinfo{author}{\bibfnamefont{D.}~\bibnamefont{Bouwmeester}},
  \bibinfo{author}{\bibfnamefont{H.}~\bibnamefont{Weinfurter}},
  \bibnamefont{and}
  \bibinfo{author}{\bibfnamefont{A.}~\bibnamefont{Zeilinger}},
\bibinfo{title}{``Experimental Entanglement Swapping: Entangling Photons That Never Interacted,"}  
\bibinfo{journal}{Phys. Rev. Lett.} \textbf{\bibinfo{volume}{80}},
  \bibinfo{pages}{3891-3894} (\bibinfo{year}{1998}).

\bibitem{hong87}
\bibinfo{author}{\bibfnamefont{C.~K.} \bibnamefont{Hong}},
  \bibinfo{author}{\bibfnamefont{Z.~Y.} \bibnamefont{Ou}}, \bibnamefont{and}
  \bibinfo{author}{\bibfnamefont{L.}~\bibnamefont{Mandel}},
\bibinfo{title}{``Measurement of subpicosecond time intervals between two photons by interference,"}
  \bibinfo{journal}{Phys. Rev. Lett.} \textbf{\bibinfo{volume}{59}},
  \bibinfo{pages}{2044-2046} (\bibinfo{year}{1987}).

\bibitem{mosley08}
\bibinfo{author}{\bibfnamefont{P.~ J.}~\bibnamefont{Mosley}},
  \bibinfo{author}{\bibfnamefont{J.~ S.}~\bibnamefont{Lundeen}},
  \bibinfo{author}{\bibfnamefont{B.~ J.}~\bibnamefont{Smith}},
  \bibinfo{author}{\bibfnamefont{P.}~\bibnamefont{Wasylczyk}},
  \bibinfo{author}{\bibfnamefont{A.~ B.}~\bibnamefont{U'Ren}}, 
  \bibinfo{author}{\bibfnamefont{C.}~\bibnamefont{Silberhorn}}, 
\bibnamefont{and}
  \bibinfo{author}{\bibfnamefont{I.~ A.}~\bibnamefont{Walmsley}},
\bibinfo{title}{``Heralded Generation of Ultrafast Single Photons in Pure Quantum States,"}
  \bibinfo{journal}{Phys. Rev. Lett.} \textbf{\bibinfo{volume}{100}},
\bibinfo{pages}{133601}
  (\bibinfo{year}{2008}).
\bibitem{cohen08} \bibinfo{author}{\bibfnamefont{O.}~\bibnamefont{Cohen}}, \bibinfo{author}{\bibfnamefont{J.S.}~\bibnamefont{Lundeen}}, \bibinfo{author}{\bibfnamefont{B.~J.} \bibnamefont{Smith}}, \bibinfo{author}{\bibfnamefont{G.}~\bibnamefont{Puentes}}, \bibinfo{author}{\bibfnamefont{P.~J.} \bibnamefont{Mosley}}, \bibnamefont{and} \bibinfo{author}{\bibfnamefont{I.~A.} \bibnamefont{Walmsley}}, \bibinfo{title}{``Tailored photon-pair generation in optical fibers,"} \bibinfo{journal}{Phys. Rev. Lett.} \textbf{\bibinfo{volume}{102}}, \bibinfo{pages}{123603} (\bibinfo{year}{2009}). 

\bibitem{garay07} \bibinfo{author}{\bibfnamefont{K.}~\bibnamefont{Garay-Palmett}}, \bibinfo{author}{\bibfnamefont{H.~J.} \bibnamefont{McGuinness}}, \bibinfo{author}{\bibfnamefont{O.}~\bibnamefont{Cohen}}, \bibinfo{author}{\bibfnamefont{J.~S.} \bibnamefont{Lundeen}}, \bibinfo{author}{\bibfnamefont{R.}~\bibnamefont{Rangel-Rojo}}, \bibinfo{author}{\bibfnamefont{A.~B.} \bibnamefont{U'ren}}, \bibinfo{author}{\bibfnamefont{M.~G.} \bibnamefont{Raymer}}, \bibinfo{author}{\bibfnamefont{C.~J.} \bibnamefont{McKinstrie}}, \bibinfo{author}{\bibfnamefont{S.}~\bibnamefont{Radic}}, \bibnamefont{and} \bibinfo{author}{\bibfnamefont{I.~A.} \bibnamefont{Walmsley}}, 
\bibinfo{title}{``Photon pair-state preparation with tailored spectral properties by spontaneous four-wave mixing in photonic-crystal fiber,"}
\bibinfo{journal}{Opt. Express} \textbf{\bibinfo{volume}{15}}, \bibinfo{pages}{14870-14886} (\bibinfo{year}{2007}).  


\bibitem{halder09} \bibinfo{author}{\bibfnamefont{M.}~\bibnamefont{Halder}}, \bibinfo{author}{\bibfnamefont{J.}~\bibnamefont{Fulconis}}, \bibinfo{author}{\bibfnamefont{B.} \bibnamefont{Cemlyn}}, \bibinfo{author}{\bibfnamefont{A.} \bibnamefont{Clark}}, \bibinfo{author}{\bibfnamefont{C.} \bibnamefont{Xiong}}, \bibinfo{author}{\bibfnamefont{W.~J.} \bibnamefont{Wadsworth}}, \bibnamefont{and} \bibinfo{author}{\bibfnamefont{J.~G.}~\bibnamefont{Rarity}}, \bibinfo{title}{``Nonclassical 2-photon interference with separated intrinsically narrowband fibre sources,"} \bibinfo{journal}{Opt. Express} \textbf{\bibinfo{volume}{17}}, \bibinfo{pages}{4670-4676} (\bibinfo{year}{2009}).

\bibitem{soller09} \bibinfo{author}{\bibfnamefont{C.}~\bibnamefont{Soller}}, \bibinfo{author}{\bibfnamefont{B.}~\bibnamefont{Brecht}}, \bibinfo{author}{\bibfnamefont{P.}~\bibfnamefont{J.}~\bibnamefont{Mosley}}, \bibinfo{author}{\bibfnamefont{Z.} \bibnamefont{Leyun}}, \bibinfo{author}{\bibfnamefont{A.} \bibnamefont{Podlipensky}}, \bibinfo{author}{\bibfnamefont{N.~Y.} \bibnamefont{Joly}}, \bibinfo{author}{\bibfnamefont{P.~St.~J.}~\bibnamefont{Russell}}, \bibnamefont{and} \bibinfo{author}{\bibfnamefont{C.}~\bibnamefont{Silberhorn}}, \bibinfo{title}{``Bridging Visible and Telecom Wavelengths with a Single-Mode Broadband Photon Pair Source,"} \bibinfo{journal}{arXiv:0908.2932} (\bibinfo{year}{2009}).


\bibitem{heersink05}
\bibinfo{author}{\bibfnamefont{J.}~\bibnamefont{Heersink}},
  \bibinfo{author}{\bibfnamefont{V.}~\bibnamefont{Josse}},
  \bibinfo{author}{\bibfnamefont{G.}~\bibnamefont{Leuchs}}, \bibnamefont{and}
  \bibinfo{author}{\bibfnamefont{U.~L.} \bibnamefont{Andersen}},
\bibinfo{title}{``Efficient polarization squeezing in optical fibers,"}
  \bibinfo{journal}{Opt. Lett.} \textbf{\bibinfo{volume}{30}},
  \bibinfo{pages}{1192-1194} (\bibinfo{year}{2005}).

\bibitem{milanovic09} \bibinfo{author}{\bibfnamefont{J.}~\bibnamefont{Milanovic}}, \bibinfo{author}{\bibfnamefont{M.}~\bibnamefont{Lassen}}, \bibinfo{author}{\bibfnamefont{U.~L.} \bibnamefont{Andersen}}, \bibnamefont{and} \bibinfo{author}{\bibfnamefont{G.}~\bibnamefont{Leuchs}}, 
\bibinfo{title}{``A Novel Method for Polarization Squeezing with Photonic Crystal Fibers,"}
\bibinfo{journal}{arXiv:0902.4597v1}  (\bibinfo{year}{2009}).


\end{thebibliography}
\end{document}